\documentstyle[11pt,preprint,prd,aps]{revtex}

\tighten
\begin{document}

\draft

\title{The Bottom Mass Prediction\\ in Supersymmetric Grand Unification:\\
  Uncertainties and Constraints}

\author{Paul Langacker and Nir Polonsky}
\address{Department of Physics, University of Pennsylvania\\
Philadelphia, Pennsylvania, 19104, USA\\}

\date{May 1993, UPR-0556T}

\maketitle

\begin{abstract}
Grand unified theories
often predict unification of Yukawa couplings
(e.g., $h_{b} = h_{\tau}$), and thus
certain relations among fermion masses.
The latter can
distinguish these
from models that predict only coupling constant unification.
The implications of
Yukawa couplings of the heavy-family
in the supersymmetric extension of the standard model
(when embedded in a GUT) are discussed.
In particular, uncertainties associated with $m_{t}$ and $m_{b}$,
threshold corrections at the low-scale, and threshold and
nonrenormalizable-operator
corrections associated with a grand-unified
sector at the high-scale are parametrized and estimated.
The implication of these and of the correlation between
$m_{t}$ and the prediction for $\alpha_{s}$  are discussed.
Constraints on the $\tan\beta$ range
in such models and an upper bound on the $t$-quark
pole mass are given and are shown to be affected by the
$\alpha_{s}-m_{t}$ correlation. Constraints on the low-scale
thresholds are found to be weakened by uncertainties associated with
the high-scale.
\end{abstract}

\newcommand{\AdBF}{U. Amaldi, W. de Boer and H. F\"{u}rstenau,
Phys. Lett. {\bf B260}, 447 (1991)}
\newcommand{\ALS}{B. Ananthanarayan, G. Lazarides and Q. Shafi,
Phys. Rev. {\bf D44}, 1613 (1991)}
\newcommand{\ACPZ}{F. Anselmo, L. Cifarelli, A. Peterman and A. Zichichi,
Nouvo Cimento {\bf 104A}, 1817 (1991)}
\newcommand{\AKT}{I. Antoniadis, C. Kounnas and K. Tamvakis,
Phys. Lett. {\bf B119}, 377 (1982)}
\newcommand{\Arasonetal}{H. Arason et al.,
Phys. Rev. {\bf D46}, 3945 (1992)}
\newcommand{\Arasonunif}{H. Arason et al., Phys. Rev. Lett. {\bf 67},
2933 (1991)}
\newcommand{\Arasonckm}{H. Arason, D. J. Casta\~{n}o, E. J. Piard
and P. Ramond,
Phys. Rev. {\bf D47}, 232 (1993)}
\newcommand{\ANproton}{R. Arnowitt and P. Nath, Phys. Rev. Lett. {\bf 69},
725 (1992); Phys. Rev. {\bf D46}, 3981 (1992)}
\newcommand{\BM}{K. S. Babu and R. N. Mohapatra, Bartol preprint
BA-92-54 (1992)}
\newcommand{\BKMN}{M. Bando, T. Kugo, N. Maekawa and H. Nakano,
Mod. Phys. Lett. {\bf A7}, 3379 (1992)}
\newcommand{\Barbierirev}{R. Barbieri, Riv. Nuovo Cimento {\bf 11},
1 (1988)}
\newcommand{\BHall}{R. Barbieri and L. J. Hall, Phys. Rev. Lett. {\bf 68},
752 (1992)}
\newcommand{\BBO}{V. Barger, M. S. Berger and P. Ohmann,
Phys. Rev. {\bf D47}, 1093 (1993);
V. Barger, M. S. Berger, P. Ohmann and R. J. N. Philips,
Wisconsin preprint MAD/PH/755 (1993)}
\newcommand{\BJ}{J. E. Bj\"{o}rkman and D. R. T. Jones,
Nucl. Phys. {\bf B259}, 533 (1985)}
\newcommand{\BEGN}{A. J. Buras, J. Ellis, M. K. Gaillard and
D. V. Nanopoulos, {\it ibid.} {\bf 135}, 66 (1978)}
\newcommand{\CJvN}{D. M. Capper, D. R. T. Jones and
P. van Nieuwenhuizen, Nucl. Phys. {\bf B167}, 479 (1980)}
\newcommand{\CHAK}{P. H. Chankowski, Phys. Rev. {\bf D41}, 2877 (1990)}
\newcommand{\CEG}{M. S. Chanowitz, J. Ellis and M. K. Gaillard,
Nucl. Phys. {\bf B128}, 506 (1977)}
\newcommand{\CPW}{M. Carena, S. Pokorski and C. E. M. Wagner,
Max-Planck preprint MPI-Ph/93-10 (1993)}
\newcommand{\DFSirlin}{G. Degrassi, S. Fanchiotti and A. Sirlin,
Nucl. Phys. {\bf B351}, 49 (1991)}
\newcommand{\DG}{S. Dimopoulos and H. Georgi, Nucl. Phys. {\bf B193},
150 (1981)}
\newcommand{\DN}{M. Drees and M. M. Nojiri, Nucl. Phys. {\bf B369}, 54 (1992)}
\newcommand{\Drees}{J. Ellis et al., Phys. Lett. {\bf B155}, 381 (1985);
M. Drees, {\it ibid.} {\bf 158}, 409 (1985)}
\newcommand{\EJ}{M. B. Einhorn and D. R. T. Jones,
{\it ibid.} {\bf 196}, 475 (1982)}
\newcommand{\EG}{J. Ellis and M. K. Gaillard, Phys. Lett.
{\bf B88}, 315 (1975)}
\newcommand{\EKN}{J. Ellis, S. Kelley and D. V. Nanopoulos,
Phys. Lett. {\bf B249}, 441 (1990)}
\newcommand{\FLY}{P. H. Frampton, J. T. Liu and M. Yamaguchi,
Phys. Lett. {\bf B277}, 130 (1992)}
\newcommand{\Fritz}{H. Fritzsch, Phys. Lett. {\bf B70}, 436 (1977);
{\it ibid.} {\bf 73}, 317 (1978);
Nucl. Phys. {\bf B155}, 189 (1979)}
\newcommand{\GL}{J. Gasser and H. Leutwyler,
Phys. Rep. {\bf C87}, 77 (1982)}
\newcommand{\GG}{H. Georgi and S. L. Glashow,
Phys. Rev. Lett. {\bf 32}, 438 (1974)}
\newcommand{\GQW}{H. Georgi, H. R. Quinn and S. Weinberg,
{\it ibid.} {\bf 33}, 451 (1974)}
\newcommand{\GJ}{H. Georgi and C. Jarlskog, Phys. Lett. {\bf B86}, 297 (1979)}
\newcommand{\GHallS}{A. Giveon, L. J. Hall and U. Sarid,
Phys. Lett. {\bf B271}, 138 (1991)}
\newcommand{\GBGS}{N. Gray, D. J. Broadhurst, W. Grafe and K. Schilcher,
Z. Phys. {\bf C48}, 673 (1990)}
\newcommand{\HKrev}{H. E. Haber and G. L. Kane,
Phys. Rep. {\bf C117}, 75 (1985)}
\newcommand{\Hall}{L. Hall, Nucl. Phys. {\bf B178}, 75 (1981)}
\newcommand{\HallS}{L. J. Hall and U. Sarid, LBL preprint LBL-32905 (1992)}
\newcommand{\HillNRO}{C. T. Hill, Phys. Lett. {\bf B135}, 47 (1984)}
\newcommand{\Hillfixed}{C. T. Hill, Phys. Rev. {\bf D24}, 691 (1981)}
\newcommand{\HMY}{J. Hisano, H. Murayama and T. Yanagida,
Phys. Rev. Lett. {\bf 69}, 1014 (1992);
Tohoku preprint TU-400 (1992)}
\newcommand{\IL}{L. E. Ib\'{a}\~{n}ez and C. L\'{o}pez,
Nucl. Phys., {\bf B233}, 511 (1984)}
\newcommand{\IR}{L. E. Ib\'{a}\~{n}ez and G. G. Ross,
Nucl. Phys. {\bf B368}, 3 (1992)}
\newcommand{\IRrev}{L. E. Ib\'{a}\~{n}ez and G. G. Ross,
SLAC preprint PUB-6412/92
(1992) (in {\it Perspectives in Higgs Physics}, ed. G. L. Kane,
World Scientific, Singapore, in press)}
\newcommand{\KLN}{S. Kelley, J. L. Lopez, and D. V. Nanopoulos,
Phys. Lett. {\bf B274}, 387 (1992)}
\newcommand{\PGLLUO}{P. Langacker and M. Luo, Phys. Rev. {\bf D44}, 817 (1991)}
\newcommand{\PGLNIR}{P. Langacker and N. Polonsky,
Phys. Rev. {\bf D47}, 4028 (1993)}
\newcommand{\PGLproton}{P. Langacker, Pennsylvania preprint UPR-0539T (1992)}
\newcommand{\PGLtasi}{P. Langacker, Pennsylvania preprint UPR-0555T (1993)}
\newcommand{\PGLrev}{P. Langacker, Phys. Rep. {\bf C72}, 185 (1981);
in {\it Proceedings of the
Ninth Workshop on Grand Unification},
Aix les Bains, France, 1988, ed. R. Barloutaud
(World Scientific, Singapore, 1988) p.3}
\newcommand{\MOHbook}{R. N. Mohapatra, {\it Unification and Supersymmetry}
(Springer, New York, 1986, 1992)}
\newcommand{\Naculich}{S. G. Naculich,
Johns Hopkins preprint JHU-TIPAC-930002 (1993)}
\newcommand{\Narison}{S. Narison, Phys. Lett. {\bf B197}, 405 (1987);
{\it ibid.} {\bf 216}, 191 (1989)}
\newcommand{\Nillesrev}{H. P. Nilles, Phys. Rep. {\bf C110}, 1 (1984);
{\it Testing The Standard Model}, ed. M. Cveti\v{c} and P. Langacker
(World Scientific, Singapore, 1991) p. 633}
\newcommand{\OP}{M. Olechowski and S. Pokorski, Max-Planck preprint
MPI-Ph/92-118 (1992)}
\newcommand{\Rossbook}{G. G. Ross, {\it Grand Unified Theories}
(Benjamin, New York, 1984)}
\newcommand{\RossP}{B. Pendleton and G. G. Ross, Phys. Lett.
{\bf B98}, 291 (1981)}
\newcommand{\RossR}{G. G. Ross and R. G. Roberts,
Nucl. Phys. {\bf B377}, 571 (1992).
More recently, the Ross and Roberts  analysis was extended to include
dark-matter
considerations by R. G. Roberts and L. Roszkowski,
Michigan preprint UM-TH-33/92 (1992)}
\newcommand{\SakaiSUSY}{N. Sakai, Z. Phys. {\bf C11}, 153 (1981)}
\newcommand{\Siegel}{W. Siegel, Phys, Lett. {\bf B84}, 193 (1979)}
\newcommand{\SW}{Q. Shafi and C. Wetterich, Phys. Rev. Lett. {\bf 52},
875 (1984)}
\newcommand{\WittenSUSY}{E. Witten, Nucl. Phys. {\bf B188}, 513 (1981)}
\newcommand{\ALPHA}{\DFSirlin}
\newcommand{\ALPHAS}{S. Bethke and S. Catani, CERN preprint TH.6484/92 (1992)}
\newcommand{\LEP}{LEP collaborations, Phys. Lett. {\bf B276}, 247 (1992)}
\newcommand{\MTAU}{H. Marsiske, SLAC preprint PUB-5977 (1992)}
\newcommand{\DRbar}{\Siegel; \CJvN; \AKT.}
\newcommand{\GUT}{\GG; \GQW. For reviews, see \PGLrev; \Rossbook;
\MOHbook. }
\newcommand{\MBMTAU}{\CEG; \BEGN.}
\newcommand{\MSbar}{W. A. Bardeen, A. J. Buras, D. W. Duke and T. Muta,
Phys. Rev. {\bf D18}, 3998 (1978) and references therein. For the weak angle
definition, see A. Sirlin, Phys. Lett. {\bf B232}, 123 (1989); S. Fanchiotti
and A. Sirlin, Phys. Rev. {\bf D41}, 319 (1990); A. Sirlin,
Nucl. Phys. {\bf B332}, 20 (1990); and references therein.}
\newcommand{\MSSM}{\DG; \SakaiSUSY; \WittenSUSY. For reviews, see
\Nillesrev; \HKrev; \Barbierirev; \IRrev.}
\newcommand{\MSSMunifnew}{\EKN; \PGLLUO; \AdBF; \ACPZ.}
\newcommand{\TEXTURES}{\Fritz; \GJ.}
\newcommand{\DHR}{S. Dimopoulos, L. J. Hall and S. Raby,
Phys. Rev. Lett. {\bf 68}, 1984 (1992); Phys. Rev. {\bf D45}, 4192 (1992);
{\it ibid} {\bf 46}, 4793 (1992);
{\it ibid} {\bf 47}, R3697 (1993);
V. Barger, M. S. Berger, T. Han and M. Zralek,
Phys. Rev. Lett. {\bf 68}, 3394 (1992);
G. W. Anderson, S. Rabi, S. Dimopoulos and L. J. Hall,
Phys. Rev. {\bf D47}, R3702 (1993).}
\newcommand{\FTEXTURE}{K. S. Babu and Q. Shafi, Bartol preprints
BA-92-27 (1992); BA-92-70 (1992); BA-93-09 (1993).}
\newcommand{\GTEXTURE}{G. Giudice, Mod. Phys. Lett. {\bf A7}, 2429 (1992);
H. Dreiner, G. K. Leontaris and N. D. Tracas, National Technical
University preprint NTUA 33/92 (1992);
G. K. Leontaris and N. D. Tracas, NTUA 37/92 (1992).}
\newcommand{\Yukawaunif}{\ALS; \BKMN; \Arasonunif; \GHallS; \KLN.
For discussion of the non-supersymmetric case, see, for example,
\FLY.}
\newcommand{\alphag}{\mbox{$\alpha_{G}$} }
\newcommand{\alphagns}{\mbox{$\alpha_{G}$}}
\newcommand{\als}{\mbox{$\alpha_{s}(M_{Z})$} }
\newcommand{\alsns}{\mbox{$\alpha_{s}(M_{Z})$}}
\newcommand{\alp}{\mbox{$\alpha(M_{Z})$} }
\newcommand{\alpns}{\mbox{$\alpha(M_{Z})$}}
\newcommand{\sth}{\mbox{$s^{2}(M_{Z})$} }
\newcommand{\sthns}{\mbox{$s^{2}(M_{Z})$}}
\newcommand{\mg}{\mbox{$M_{G}$} }
\newcommand{\mgns}{\mbox{$M_{G}$}}
\newcommand{\hb}{\mbox{$h_{b}$} }
\newcommand{\hbns}{\mbox{$h_{b}$}}
\newcommand{\hbg}{\mbox{$h_{b}(M_{G})$} }
\newcommand{\hbgns}{\mbox{$h_{b}(M_{G})$}}
\newcommand{\htau}{\mbox{$h_{\tau}$} }
\newcommand{\htauns}{\mbox{$h_{\tau}$}}
\newcommand{\htaug}{\mbox{$h_{\tau}(M_{G})$} }
\newcommand{\htaugns}{\mbox{$h_{\tau}(M_{G})$}}
\newcommand{\htop}{\mbox{$h_{t}$} }
\newcommand{\htopns}{\mbox{$h_{t}$}}
\newcommand{\htg}{\mbox{$h_{t}(M_{G})$} }
\newcommand{\htgns}{\mbox{$h_{t}(M_{G})$}}
\newcommand{\ai}{({\it i}) }
\newcommand{\ains}{({\it i})}
\newcommand{\aii}{({\it ii}) }
\newcommand{\aiins}{({\it ii})}
\newcommand{\mt}{\mbox{$m_{t}$} }
\newcommand{\mtns}{\mbox{$m_{t}$}}
\newcommand{\mtp}{\mbox{$m_{t}^{pole}$} }
\newcommand{\mtpns}{\mbox{$m_{t}^{pole}$}}
\newcommand{\mtau}{\mbox{$m_{\tau}$} }
\newcommand{\mtauns}{\mbox{$m_{\tau}$}}
\newcommand{\mtaup}{\mbox{$m_{\tau}^{pole}$} }
\newcommand{\mtaupns}{\mbox{$m_{\tau}^{pole}$}}
\newcommand{\mb}{\mbox{$m_{b}$} }
\newcommand{\mbns}{\mbox{$m_{b}$}}
\newcommand{\mbo}{\mbox{$m_{b}^{0}$} }
\newcommand{\mbons}{\mbox{$m_{b}^{0}$}}
\newcommand{\mbp}{\mbox{$m_{b}^{pole}$} }
\newcommand{\mbpns}{\mbox{$m_{b}^{pole}$}}
\newcommand{\tgb}{\mbox{$\tan \beta$} }
\newcommand{\tgbns}{\mbox{$\tan \beta$}}
\newcommand{\mz}{\mbox{$M_{Z}$} }
\newcommand{\mzns}{\mbox{$M_{Z}$}}
\newcommand{\gev}{\mbox{ GeV} }
\newcommand{\tev}{\mbox{ TeV} }
\newcommand{\gevns}{\mbox{GeV}}
\newcommand{\tevns}{\mbox{TeV}}
\newcommand{\fors}{\mbox{ for $i$ = 1,\, 2,\, 3,} }
\newcommand{\forss}{\mbox{ for $i$ = 1,\, 2,\, 3.} }
\newcommand{\for}{\mbox{ for $i$ = 1,2,3,} }
\newcommand{\fixed}{\mbox{fixed} }
\newcommand{\ms}{\mbox{$\overline{MS}$} }
\newcommand{\msns}{\mbox{$\overline{MS}$}}
\newcommand{\ro}{\mbox{$\rho^{^{-1}}$} }
\newcommand{\rons}{\mbox{$\rho^{^{-1}}$}}
\newcommand{\roals}{\mbox{$\rho^{^{-1}}_{\alpha_{3}}$} }
\newcommand{\roalsns}{\mbox{$\rho^{^{-1}}_{\alpha_{3}}$}}
\newcommand{\roeff}{\mbox{$\rho^{^{-1}}_{eff}$} }
\newcommand{\roeffns}{\mbox{$\rho^{^{-1}}_{eff}$}}
\newcommand{\rot}{\mbox{$\rho^{^{-1}}_{t}$} }
\newcommand{\rotns}{\mbox{$\rho^{^{-1}}_{t}$}}
\newcommand{\rop}{\mbox{$\rho^{^{-1}}_{_{F}}$} }
\newcommand{\ropns}{\mbox{$\rho^{^{-1}}_{_{F}}$}}
\newcommand{\rofix}{\mbox{$\rho^{^{-1}}_{fixed}$} }
\newcommand{\rofixns}{\mbox{$\rho^{^{-1}}_{fixed}$}}
\newcommand{\rosusy}{\mbox{$\rho^{^{-1}}_{_{Z}}$} }
\newcommand{\rosusyns}{\mbox{$\rho^{^{-1}}_{_{Z}}$}}
\newcommand{\roheavy}{\mbox{$\rho^{^{-1}}_{_{G}}$} }
\newcommand{\roheavyns}{\mbox{$\rho^{^{-1}}_{_{G}}$}}
\newcommand{\ronro}{\mbox{$\rho^{^{-1}}_{_{NRO}}$} }
\newcommand{\ronrons}{\mbox{$\rho^{^{-1}}_{_{NRO}}$}}

\newpage
\section{Introduction}
\label{sec:intro}
Recent LEP \cite{lep} and other precision electroweak data
is known \cite{9091} to be
consistent with coupling constant
unification within the minimal supersymmetric standard model (MSSM)
\cite{mssm}, in which the standard model (SM) matter is minimally extended,
i.e., the Higgs sector contains one pair of Higgs doublets and there
is a grand desert (up to small perturbations) between
the weak (low) and unification (high) scales.
Recently, it was further shown \cite{us} that  corrections associated with
the $t$-quark and Higgs scalar thresholds,
sparticle
spectrum (for example, see Ref. \cite{rr}),
Yukawa couplings, a possible embedding of the MSSM in a grand
unified theory (GUT) \cite{gut}, and nonrenormalizable effects \cite{nro},
as well as
constraints \cite{an,yan} from  proton decay non-observation \cite{proton},
introduce theoretical uncertainties but do not alter the successful
unification;
e.g., the prediction of $\als \approx 0.125 \pm 0.010$ \cite{us} agrees well
with the observed value.
Such uncertainties depend on seven different effective parameters
in addition to the $t$-quark mass and Yukawa coupling.
(The $\pm 0.010$ is
a sum (in quadrature) of the different theoretical
uncertainties estimated using reasonable
ranges for the various parameters.)
This theoretical uncertainty is sufficiently large that few
meaningful constraints
can be derived from the \als prediction by itself.
Similar conclusions
were reached by Barbieri, Hall and Sarid \cite{hall}.

 If, indeed, coupling constant unification is a hint
for a supersymmetric (SUSY)
GUT, then a next step
is to study the predicted relationships among fermion masses
in such theories \cite{ceg},
in a way that consistently incorporates the
different theoretical uncertainties listed above.
(The nature of the theoretical corrections,
and in particular the presence of adjoint representations,
also distinguishes such models from many
string-inspired ones.)
Let us assume in the following
(in addition to the MSSM)
that we have
({\it i}) Coupling constant unification, and
({\it ii}) Third-family two-Yukawa unification.
That is, at the unification point
\mg (the point above which all the GUT gauge group
supermultiplets are complete) we have $\hbg = \htaug$,
as is the case \cite{ceg}
in a minimal $SU(5)$ unification, which
we will assume below for definiteness,
and in similar unification schemes\footnote{This holds in models
such as $SU(5)$,
$SO(10)$ and $E_{6}$ for the Yukawa coupling of Higgs fields in the
fundamental ($\underline{5}$, $\underline{10}$, $\underline{27}$)
representations.}.
$h_{\alpha}$ is the MSSM Yukawa coupling of a fermion of type $\alpha$ and
$M_{G} \approx 10^{16} - 10^{17}$ GeV.

 Assumption \aii
can be incorporated into  more ambitious attempts \cite{fm1}
to explain the origin of all fermion masses. Such models,
which assume extended high-scale
structures (``textures''), were shown recently
\cite{hall2,babu,giudice,acpp,bbo,naul,mohap}
to have successful predictions as well as possible implications
for neutrino masses.
However, limiting our analysis to assumptions \ai and \aiins,
we neglect hereafter the Yukawa couplings of the first two families
(where empirically $\frac{m_{d}}{m_{s}} \approx 10\frac{m_{e}}{m_{\mu}}$,
rather than
$\frac{m_{d}}{m_{s}} \approx \frac{m_{e}}{m_{\mu}}$; the latter would be
implied
by extending assumption \aii to the first two families and
their negligibly small Yukawa couplings)
and also flavor mixings.
The usual argument goes that some perturbation modifies the
couplings or masses
of the first two families\footnote{In the texture models
mentioned above, such a mechanism is realized by
introducing large Higgs representations
(e.g., $\underline{45}$ of $SU(5)$ or $\underline{126}$
and $\underline{210}$
of $SO(10)$) and (in most cases) a set of flavor symmetries.
For a different possibility involving nonrenormalizable operators, see Ref.
\cite{eg}.}
without significantly altering \aiins.
We do not elaborate on any
such mechanism. A special case
of \aii is a third-family three-Yukawa unification, i.e.,
$\htg = \hbg = \htaug$, which is the situation in some $SO(10)$ models
involving a single complex Higgs $\underline{10}$-plet. We will
consider such a possibility as well.

 Let us stress that we do not take \aii
to be independent of \ains. The coupling constant unification assumption
by itself is not enough to
significantly constrain the MSSM parameter space.
Here, we examine whether more can be said when imposing
\aii as an additional assumption that can possibly
distinguish GUT models from some GUT-like string-inspired models
(where \aii is not expected to hold in general).
Assumption \aii was considered recently by
several groups. Some \cite{yukawaunif},
either  ({\it a}) carried out a one-loop analysis,
({\it b}) assumed a low \als (e.g., $\als \sim 0.11$,
which is lower than the value expected from coupling constant unification)
as an input,
({\it c}) ignored the correlation between \mt and the predicted value of
\alsns,
and/or ({\it d}) allowed the running $b$-quark mass, $m_{b}$, to be as high as
5 GeV
(which, as we discuss
below, is a more appropriate upper bound on the pole mass).
More recent results of
two-loop analyses  \cite{rr,bbo,cpw}
imply
a very constrained parameter space, i.e., only a small allowed area in the
$\mtp - \tgb$ plane, where \mtp is the $t$-quark pole mass and \tgb is the
ratio
of the two Higgs doublet expectation values,
$\nu_{_{h_{up}}}/\nu_{_{h_{down}}}$.
Therefore, one would hope that linking \ai with \aii
(and considering uncertainties associated with $m_{t}$ and $m_{b}$)
will result in some
useful constraints on the MSSM parameters, assuming a minimal $SU(5)$-type
unification (for example,  see Ref. \cite{rr}).

 Below, we carry out a careful analysis under
the above assumptions and consider various
theoretical uncertainties in the calculation.
We find that requiring \ai
and predicting
\als as a function of \mtp and of \tgb \cite{us}
in the range of $\sim 0.12 - 0.13$ (see Figure 1),
constrains the \tgb range allowed by \aii
more severely than suggested by previous
analyses. On the other hand,
various theoretical uncertainties can relax the constraints.
We also obtain $\sim 215$ GeV for the upper bound on \mtp
(where $\alpha_{s} - m_{t}$ correlations were taken into account).
Some information about the low-scale mass parameters
can be extracted.
However,
corrections associated with the high-scale
contribute
significantly
to the theoretical uncertainties and weaken any constraints.
The only spectrum parameter that is  strongly constrained
is \tgbns.
In agreement with other
authors, we find low- ($\sim 0.6 - 3$) and high- ($\sim 40 - 58$) \tgb allowed
regions (branches).
The former saturates the $h_{t}$ infra-red fixed-point
\cite{fixedpoint} line (the divergence line).
The $\alpha_{s} - m_{t}$ correlation
modifies the fixed-point value for $h_{t}$ and
diminishes the dependence  of the
allowed \tgb range on
$\mtp \lesssim  215$ GeV. Theoretical uncertainties
(and in particular, those associated with the high-scale) determine
the width of each branch
and, thus, the separation between the two branches.

The various data (and in particular, the $b$-quark mass) and the procedure
are reviewed in section \ref{sec:data}.
The constraints on the $\mtp - \tgb$ plane and the role of the strong coupling
are presented and discussed in section \ref{sec:plane}.
The different correction terms are
described and evaluated in greater detail in section \ref{sec:corrections}.
We summarize our conclusions in section \ref{sec:con}.
Throughout this work, we keep the philosophy
(and where relevant, the notation) we
introduced previously \cite{us}.

\section{Input Data and Procedure}
\label{sec:data}

 At the $Z$-pole,
\begin{equation}
\mz = 91.187 \pm 0.007 \,\, \gev,
\label{zi}
\end{equation}
and using the modified minimal subtraction scheme ($\overline{MS}$) \cite{ms}
the weak angle and
couplings\footnote{A predicted \als slightly above $0.13$, as it is for
a heavy enough $t$-quark (see Figure 1), does not contradict (\ref{alphas});
the \als prediction still has a fairly large theoretical
uncertainty of $\sim \pm 0.008$.} are
\cite{us,sir,bc}
\begin{equation}
\sth = 0.2324 -1.03\times 10^{-7}(\mtp(\gevns)^{2} - (138)^{2}) \pm 0.0003,
\label{sin}
\end{equation}
\begin{equation}
\frac{1}{\alp} = 127.9 \pm 0.1,
\label{alpha}
\end{equation}
\begin{equation}
\als = 0.12 \pm 0.01,
\label{alphas}
\end{equation}
where we displayed explicitly the quadratic dependence of \sth on \mtpns,
which is decoupled
from the $0.0003$ uncertainty \cite{us}. The $U(1)_{\frac{Y}{2}}$, $SU(2)_{L}$,
and $SU(3)_{c}$ couplings are given by
$\frac{1}{\alpha_{1}(m_{Z})} = \frac{3}{5} \frac{1 -
s^{2}(M_{Z})}{\alpha(M_{Z})}$,
$\frac{1}{\alpha_{2}(M_{Z})} =  \frac{s^{2}(M_{Z})}{\alpha(M_{Z})}$, and
$\frac{1}{\alpha_{3}(M_{Z})} =  \frac{1}{\alpha_{s}(M_{Z})}$,
respectively.

 For the fermion masses,
from electroweak precision data\footnote{ Slightly more recent data
yields \cite{pglnew}
$m_{t}^{pole} = 134^{+ 23}_{-28} \pm 5$ GeV
(for $m_{h^{0}} \sim 60 - 150$ GeV
and including two-loop $\alpha \alpha_{s} m_{t}^{2}$ corrections)
and $s^{2}(M_{Z}) = 0.2326 \pm 0.0006$ ($m_{t}$ free).
For our present purposes the difference with (\ref{sin}) and (\ref{mtop}) is
negligible.
The new data will be incorporated in future analyses \cite{usprog}.}
we have for the $t$-quark
\cite{us}
\begin{equation}
\mtp = 138^{+ 20}_{-25} \pm 5\,\, \gev
\label{mtop}
\end{equation}
for a Higgs mass in the range $50 - 150$ GeV,
which is appropriate for the MSSM.
The pole mass  is related to the
$\overline{MS}$ running mass, \mtns,
to leading order\footnote{
The next-to-leading correction \cite{threeloop,florida}
is $\sim 2\%$ (depending on
$\alpha_{s}$). The leading correction given here is $\sim 5\%$.
(See also the discussion of ${\mb}/{\mbp}$ below.)
We neglect the former (together with other
subleading \mt and \mtp effects) while keeping the latter.} in $\alpha_{s}$
by $\mt = (1 - \frac{4}{3}\frac{\alpha_{s}}{\pi})\mtp$.
The $\tau$-lepton ($\overline{MS}$ running) mass \cite{florida}
is given at the $Z$-pole by
\begin{equation}
m_{\tau}(\mz) = 1.7486 \pm 0.0006 \,\, \gev,
\label{mtau}
\end{equation}
which corresponds to (\ref{alpha}) and $m_{\tau}^{pole} = 1.7771 \pm 0.0005$
GeV
\cite{mtau}.

 The situation regarding the $b$-quark mass is more complicated.
There are ambiguities in the extraction of the \ms running mass
\mbns. Gasser and Leutwyler \cite{gl} point out that there is no
universal prescription
for the relevant scale where $\alpha_{s}$ is to be evaluated,
which suggests that the extraction of
\mb is to be carried out case by case, or alternatively, for a range
of $\alpha_{s}$. (We will adopt the latter.)
Gasser and Leutwyler identify (to leading order in $\alpha_{s}$)
the running mass $m_{b}(m_{b})$ with the euclidean mass parameter.
This point was emphasized by Narison, who offers an alternative
definition of $m_{b}(\mbp)$ \cite{nar}. The different
definitions introduce a scale ambiguity. Another theoretical
difficulty may arise
from the role of nonperturbative effects in the
interpretation of potential
models\footnote{The constituent mass parameter in these models
is identified with the pole mass.}
\cite{gl}. The next-to-leading correction to the ratio of the
\ms running mass to the pole mass was given more recently by Gray et al.
\cite{threeloop}, i.e.,
\begin{equation}
\mb = \mbp\left( 1 - \frac{4}{3}\frac{\alpha_{s}}{\pi}
 - 12.4(\frac{\alpha_{s}}{\pi})^{2} \right).
\label{threeloop}
\end{equation}
The above comments call for some caution, especially given
our aim of exploring the strongly
constrained $\mtp-\tgb$ plane. Let us then adopt a conservative attitude, i.e.,
\begin{equation}
m_{b}(5 \gev) \leq 4.45 \,\, \gev,
\label{mbot}
\end{equation}
which corresponds, for example, to
$\mbp \leq 5$ GeV, $\alpha_{s} \geq 0.17$, and using
(\ref{threeloop}).
The next-to-leading correction term in (\ref{threeloop}) reduces \mbns,
and $\mbp \approx 4.5$ GeV,
$\alpha_{s} \gtrsim 0.2$ implies  $\mb < 4$ GeV. For example,
$\mbp \approx 4.5$ GeV, $\alpha_{s} \approx 0.25$
gives $\mb \approx 4.0$ GeV when neglecting
the next-to-leading term, and $\mb \approx 3.7$ GeV when using
(\ref{threeloop}).
The \mb prediction, on the other hand, lies (in general) above $4$ GeV.
Given the above,
we do not specify a lower bound equivalent to (\ref{mbot}).
Also, requiring $m_{b}(4.45 \gev) \leq 4.45$ GeV (which will correspond to
$m_{b}(m_{b}) = 4.25 \pm 0.20$ GeV \cite{gl},
where we have doubled the uncertainty)
is somewhat more constraining
(e.g., $m_{b}(4.45 \gev) - m_{b}(5 \gev) \approx 0.05 - 0.15 \, \gev$ --
depending on $\alpha_{s}$).

 We use \alpns, \sthns, and the $\tau$-lepton and $t$-quark
Yukawa couplings,
\begin{equation}
h_{\tau}(\mz) = \frac{m_{\tau}(\mz)}{174\gev \times \cos \beta}
\label{htau}
\end{equation}
and
\begin{equation}
h_{t}(\mz) = \left( 1 - \frac{4}{3} \frac{\als}{\pi} \right)
\frac{\mtp}{174\gev \times \sin \beta} \, ,
\label{htop}
\end{equation}
to predict\footnote{i.e., case ({\sl b}) in the notation of Ref. \cite{us}.}
\als
and \hbns(\mzns), for a definite point in the $\mtp - \tgb$ plane.
One should note that
\htop  depends on \mtp
also via the \als correction in  (\ref{htop}) (and  via the
$\alpha_{3}$ contribution to the running -- see below).
As we pointed out, \sth
depends quadratically on \mtpns.
Therefore, we  neglect
all subleading logarithmic dependencies on \mtp
(for a discussion, see Ref. \cite{us}),
including small corrections to (\ref{htop}).
We further neglect the error bars in
(\ref{zi}) -- (\ref{alpha}) and in (\ref{mtau}).
Also, $\alpha_{i}$ are all converted
to the $\overline{DR}$ scheme, using the proper step functions \cite{dr}.
Using two-loop renormalization group equations (RGE's) \cite{bj} iteratively,
we are able to predict \als and \hbns(\mzns) as functions of \mtp and \tgbns.
We take $100 < \mtp < 200 \gev$ as
a reasonable conservative range, and constrain
\tgb only by requiring the Yukawa couplings to stay perturbative, i.e.,
$h_{\alpha}(\mu) < 3$ where $\mz < \mu < \mg$ and $\alpha = t, \, b, \, \tau$.
(This range can be also justified by requiring
two-loop contributions to the RGE's
to be less than a quarter of the one-loop ones \cite{bbo}.)
We then run down
using three-loop QCD and two-loop QED RGE's \cite{florida}
to find $\mbo(5 \gev)$, where the \mbo prediction is that of
\mbns, but without (theoretical) corrections to the RG calculation.

In any realization of the MSSM, there are
small perturbations (order of magnitude of two-loop terms)
to the grand desert and unification
assumptions, as described above. Thus, in general,
$\mb = \ro\mbo$ where
$\ro \neq 1$ is a correction parameter which incorporates the uncertainties
in the running from \mz up to \mgns.
Let us stress that in our formulation one does not
change the MSSM $\beta$-functions
to those of the SM
at \mt or at some other effective scale. Rather, leading \mtp effects are
accounted
for in (\ref{sin}), and all other such effects determine \rons.
A point in the $\mtp - \tgb$ is excluded if
 either $\htop > 3$
($\tgb \lesssim 1-2$ and/or $\mtp \gtrsim 215$ GeV), $\hb \, (\sim \htau) > 3$
($\tgb \gtrsim 58 $), or
\begin{equation}
\mb(5 \gev) = \rho^{-1} \mbo(5 \gev) > 4.45 \, \gev.
\label{ro1}
\end{equation}

Incorporating uncertainties associated with
\mtp and Yukawa couplings
(in addition to the $\overline{DR}$ conversion step functions)
in the numerical procedure\footnote{In the notation of ref. \cite{us},
$\Delta_{i}^{conversion}$, $\Delta_{i}^{top}$, and $\Delta_{i}^{Yukawa}$,
are all directly incorporated in the calculation.}
we have to further consider
uncertainties associated with the sparticle and Higgs thresholds,
high-scale thresholds, and Planck-scale
nonrenormalizable operators. For simplicity, we will
assume that we have one heavy ($M_{H} \gg \mz$)
Higgs doublet that decouples with the sparticles, and
another light ($m_{h} \sim \mz$) SM-like doublet that is responsible for all
fermion
masses\footnote{In such a case,
$SU(2)$ breaking effects are, in general, negligible above \mtns.}.
We are able to
obtain an (approximate) analytic expression for \rons
by expanding one-loop expressions
around their unperturbed values.
This will be carried out in section \ref{sec:corrections},
where we study the different contributions to \ro
in GUT models, and estimate \ro in the minimal $SU(5)$
model. High-scale corrections to the
coupling constant unification (and not the details of the sparticle spectrum)
constitute the larger uncertainty.
We  take
\begin{equation}
\ro = 1.00 \pm 0.15,
\label{ro3}
\end{equation}
which is a conservative  estimate derived for reasonable ranges of the
various correction parameters.
Using (\ref{ro3}), the exclusion condition (\ref{ro1}) reads
\begin{equation}
\mb(5 \gev) \geq 0.85\mbo(5 \gev) > 4.45 \, \gev.
\label{ro4}
\end{equation}

\section{The $\mtp - \tgb$ Plane}
\label{sec:plane}

Given the above, we find that assumptions \ai and \aii allow a low-\tgb branch
and a high-\tgb
branch.
The allowed parameter space is shown in Figure 2, where the narrow strip
corresponding
to three-Yukawa unification is also indicated. The low-\tgb
branch is shown in greater detail in
Figure 3, where the lines corresponding to $\ro = 1$  and $\htg = 2$ are
displayed
for comparison.
The former is, in fact, the $h_{t}$ infra-red fixed-point \cite{fixedpoint}
line,
which is the $h_{t}$-divergence line.
($h_{t}(M_{Z}) > h_{t}^{fixed}$ $\Rightarrow$  $h_{t}(\mu) \gg 1$ for $\mu <
\mg$.)
This point was also discussed recently in Ref. \cite{bbo,cpw}.
$\ro \neq 1$ only slightly extends the allowed low-\tgb range.
It is also interesting to note that constraints from proton decay via
dimension-five operators would
exclude the high-\tgb branch for $\ro \equiv 1$ (i.e., $\tgb \lesssim 4.7$
\cite{an}).
However, once correction terms are included, \mg can grow significantly
\cite{yan,us,hall}
and no useful constraints on \tgb can be derived from proton decay
non-observation \cite{yan}.
For comparison, we show in Figure 4 the equivalent parameter space  with
(\ref{ro4}) replaced
by $0.85\mbo(4.45 \gev) > 4.45$ GeV.
The allowed \tgb range is reduced by $\sim 0.03 - 0.10$ for the low-\tgb branch
and
by $3 - 4$ for the high-\tgb one.
Replacing (\ref{ro4}) with $0.85\mbo(\sim 5.1 \gev) \gtrsim 4.6$ GeV
would have a similar but opposite effect
(i.e., slightly decreasing the separation between the two branches).
A smaller (larger) uncertainty in (\ref{ro3}) will have an effect
similar to the former (latter). The \rons-range estimate
(and the \mb upper bound) determine the width of each branch,
and thus the excluded intermediate \tgb range.
Perturbative consistency
(i.e., the divergence lines discussed above)
excludes the very small and the very large \tgb ranges
and determines the upper bound on \mtpns, $\mtp \lesssim 215 \pm 10$ GeV,
where the $\pm 10$ GeV  uncertainty is due to $\rho_{top}$.

The $h_{t}$-divergence line eventually becomes
approximately  parallel to the \tgbns-axis (and
determines the upper bound
$\mtp \lesssim 215$ GeV).
 Some intermediate values of \tgb are thus allowed for $\mtp > 200$ GeV.
(The $h_{b}$ and $h_{t}$ divergence lines meet near the $h_{b} = h_{t}$ line.)
Otherwise,
Yukawa unification at the grand unification scale
is ruled out\footnote{For a large \tgb
($\nu_{_{h_{down}}} \ll\nu_{_{h_{up}}}, \, \nu$)
some caution may be required regarding the scale
at which the Higgs potential is minimized and $\tan\beta$
is defined,
as was pointed out by Bando et al. \cite{yukawaunif}
and by Chankowski \cite{chak}.}
if $2.7 \lesssim \tgb \lesssim 40$.
Furthermore, the low-\tgb branch, where
$\htop \sim 1$ and which
many would consider a more natural choice,
saturates the fixed-point line and has to be
adjusted to a few parts in a hundred
(a few parts in a thousand, if $\ro = 1$) for a given
\mtp (see Figure 3).
The large-\tgb branch is more spread and implies, in
general, a much lower \htop
and $\htop < \hb$.
(\htop can still be large for a large enough \mtpns, and
$\htop > \hb$ above the three-Yukawa unification strip.)
While we find no constraints on $\mtp \lesssim 215$ GeV
from two-Yukawa unification,
three-Yukawa unification is ruled out unless $169 \lesssim \mtp \lesssim 196$
GeV.
(A slightly larger range, i.e., $\mtp \gtrsim 160$ GeV, is allowed when
one includes corrections to the $h_{t}/h_{b}$ ratio, which  induce a $\sim 5\%$
theoretical
uncertainty. We comment more on this point in section \ref{sec:corrections}.)
One expects mutual implications \cite{usprog} between
the above observations and
radiative-breaking of $SU(2)\otimes U(1)$, an attractive feature of the MSSM
that prefers $\htop > \hb$ \cite{htllhb}.

 To demonstrate the effect of calculating \mb using the predicted \als
rather than a fixed input value,
i.e., of associating the Yukawa coupling unification with the rather high
values of
\als predicted by  $\alpha_{1} - \alpha_{2}$ unification,
we compare Figure 2 with Figures 5 -- 7.
There, \als is fixed ($\als = 0.11,\,0.12,\,0.13,$ in Figure 5, 6, 7,
respectively),
and thus assumption \ai is relaxed; i.e., for $\als =0.11$ ($0.12$, $0.13$)
there is a $\sim 7\%$ ($\sim 3\%$) split between
$\alpha_{3}(\mg)$ and the \alphag defined by $\alpha_{1}$ and $\alpha_{2}$.
Let us stress that the
different corrections are not treated on equal footing in this case,
because some are included in \ro while others (like NRO's) are absorbed
in the fixed value of \alsns.
Furthermore, the appropriateness of this decomposition depends
on which type of uncertainties shift the predicted \alsns.
(We elaborate more on this point in section \ref{sec:corrections}.)
Nevertheless,
the comparison illustrates that a low \als is preferred by \mbns.
The allowed parameter space for
$\als = 0.11$ (Figure 5) is much larger than that for $\als = 0.13 $ (Figure
7).
For a lower value of $\alpha_{s}$ the radiative corrections that reduce \hb are
diminished,
and thus a given $\htau(\mg) = \hb(\mg)$ implies a lower $\hb(\mz)$.
However, the low value  $\als = 0.11$ requires large
corrections to
the coupling constant unification.

The above
discussion also explains the slight differences between our results
and those of previous analyses. Requiring \ai and using (\ref{sin}) for \sth
imply that
\als grows with \mtp \cite{us}, e.g., $\als \sim 0.12$ for $\mtp \sim 100$ GeV,
and
$\als \sim 0.13$ for $\mtp \sim 180$ GeV (see Figure 1). Indeed,
Figure 2 roughly coincides with Figure 6 for the former
and with Figure 7 for the latter.
\htopns(\mzns)
in (\ref{htop}) $\sim \frac{m_{t}^{pole}}{\sin\beta}$, but is diminished by
(the \mtp dependent) \alsns.
These all affect the balance between positive
and negative contributions to the Yukawa coupling RGE's (i.e., the
fixed-points),
and thus modify the \hbns(\mzns) prediction
and increase the upper bound
on \mtpns.

\section{The Correction Terms}
\label{sec:corrections}
 We now turn to a detailed discussion of the correction parameter, \rons.
The coupling constant two-loop RGE's are solvable
analytically, and it is convenient to write \cite{ej}
\begin{equation}
\frac{1}{\alpha_{i}(M_{Z})} = \frac{1}{\alpha_{G}} + b_{i}t + \theta_{i}
+ H_{i} - \Delta_{i} \,\,\fors
\label{alphamz}
\end{equation}
where $t = \frac{1}{2\pi}\ln \frac{M_{G}}{M_{Z}} \approx 5.3$ is the relevant
scale parameter,
and $\alpha_{G} \approx \frac{1}{24}$ is the coupling constant at the
unification point, \mgns.
$b_{i} = 6.6,\,1,\,-3, \fors$ respectively,
are the one-loop $\beta$-function coefficients;
$\theta_{i} \approx 0.7,\,1.1,\,0.6, \fors$
are the two-loop corrections;
$H_{i}$ are negligible Yukawa coupling two-loop contributions;
and the functions $\Delta_{i}$ incorporate all other corrections to the
calculation of order of magnitude consistent with $\theta_{i}$.
In our scheme, $\alpha_{1}$ and $\alpha_{2}$ are inputs. By taking linear
combinations
we obtain three predictions, i.e.,
\begin{mathletters}
\label{predictions}
\begin{equation}
\als = \alpha_{s}^{0}(\mz)[\alpha_{1},\, \alpha_{2},\,\theta_{i}]
+ (\alpha_{s}^{0}(\mz))^{2}\Delta_{\alpha_{s}}[\Delta_{1},\, \Delta_{2},\,
\Delta_{3}],
\label{predals}
\end{equation}
\begin{equation}
\frac{1}{\alpha_{G}} =
\frac{1}{\alpha_{G}^{0}}[\alpha_{1},\, \alpha_{2},\,\theta_{1},\,\theta_{2}]
+ \Delta_{\alpha_{G}}[\Delta_{1},\, \Delta_{2}],
\label{predalphag}
\end{equation}
\begin{equation}
t  =
t^{0}[\alpha_{1},\, \alpha_{2},\,\theta_{1},\,\theta_{2}]
+ \Delta_{t}[\Delta_{1},\, \Delta_{2}],
\label{predti}
\end{equation}
\end{mathletters}
where we explicitly separated the two-loop predictions with no corrections
($\Delta_{i} = 0$) from the contribution of the correction functions,
$\Delta_{i}$.
The expressions for
$\Delta_{\alpha_{s}}$, $\Delta_{\alpha_{G}}$ and $\Delta_{t}$
are given in Appendix \ref{sec:app}.

The integration of the two-loop RGE's for the Yukawa couplings \cite{bj} is
rather
complicated and has to be done numerically.
To estimate the theoretical correction terms it is useful to display
the (one-loop) RGE's, i.e.,
\begin{equation}
\frac{dy_{\alpha}}{y_{\alpha}} = \left[ \sum_{i = 1,2,3} b_{\alpha;i}\alpha_{i}
+ \sum_{\beta = t,b,\tau} b_{\alpha;\beta}y_{\beta}  + ...\right]dt',
\label{rges}
\end{equation}
where $y_{\alpha} = \frac{h_{\alpha}^{2}}{4\pi} $  for $\alpha = t,\,b,
\,\tau;$
$t' = \frac{1}{2\pi}\ln{\frac{\mu'}{M_{Z}}}$; and we
have omitted higher-order terms.
$b_{b;i} = -\frac{7}{15},\, -3,\, -\frac{16}{3},\fors$ respectively; and
$b_{b;\beta} = 1,\, 6,\, 1,$ for $\beta = t,\,b, \,\tau.$
($b_{\tau;i} =  -\frac{9}{5},\,  -3,\,0;$  $b_{\tau;\beta} = 0,\, 3,\, 4;$
$b_{t;i} =  -\frac{13}{15},\,  -3,\, -\frac{16}{3};$ and $b_{t;\beta} = 6,\,
1,\, 0.$)
The balance between the negative $b_{\alpha;i}\alpha_{i}$ and the positive
$b_{\alpha;\beta}y_{\beta}$ terms determines the infra-red fixed point in the
Yukawa coupling renormalization flow \cite{fixedpoint}.
{}From (\ref{rges}) we obtain
\begin{equation}
\hb(\mg) = \hb(\mz) \times \prod_{i = 1}^{3} \left(
\frac{\alpha_{i}^{OL}(\mg)}{\alpha_{i}^{OL}(\mz)}
\right)^{\frac{b_{b;i}}{2b_{i}}} \times F_{b} \times \Theta_{b} \times
\rho_{b},
\label{hbmg}
\end{equation}
and similarly for \htauns({\mgns}).
The $\alpha_{i}^{OL}$ are the one-loop ($OL$) couplings
(i.e., $\theta_{i} = H_{i} = 0$ in (\ref{alphamz})).
Substituting instead two-loop ($TL$) (or input) expressions one has
to compensate by properly modifying the two-loop
correction, $\Theta_{b}$.
$F_{b}$ is the correction due to the non-negligible
Yukawa contribution at one-loop, i.e., $\int{b_{\alpha;\beta}y_{\beta}dt'}$.
$\rho_{b}$ incorporates the theoretical uncertainties in the RG calculation.

{}From (\ref{hbmg}) and the equivalent
expression for $h_{\tau}$ (and assuming
$h_{b}(\mgns) = h_{\tau}(\mgns)$)  we have
\begin{equation}
\mb(\mz) = \mtau(\mz)\times
\left( \frac{\alpha_{3}^{OL}(\mz)}{\alpha_{G}^{OL}} \right)^{\frac{8}{9}}
\left( \frac{\alpha_{1}^{OL}(\mz)}{\alpha_{G}^{OL}} \right)^{\frac{10}{99}}
\times
 F^{-1} \times \Theta^{-1} \times \ro,
\label{mbmz}
\end{equation}
where $F = \frac{F_{b}}{F_{\tau}}$, $\Theta =
\frac{\Theta_{b}}{\Theta_{\tau}}$,
and $\rho = \frac{\rho_{b}}{\rho_{\tau}}$. Setting $\Theta = \rho = 1$,
substituting the one-loop expressions for $\alpha_{i}$
and $\alpha_{G}$,
and assuming negligible Yukawa couplings (i.e., $F \approx 1$)
gives an exact well-known one-loop expression.
$F^{-1}$ can be estimated analytically
for $\htop \gg \hb,\,\htau$ \cite{il}, i.e.,
\begin{equation}
F^{-1} \approx (1 + 11\htop^{2}(\mg))^{-\frac{1}{12}},
\label{f}
\end{equation}
which gives $F^{-1} \sim 0.68$ for $\htopns(\mgns) \sim 3$.
In general, however, a numerical analysis is required to fully incorporate
$(F,\,\,\Theta) \neq 1$. The \mbons(\mzns) that we calculate is given  by
(\ref{mbmz})
with $\ro = 1$ and numerical values for $F$ and $\Theta$.

Before we turn to a rather technical derivation of the correction parameter
\rons,
let us discuss a simple toy model and point out the ways in which
it gets complicated. If the ideal desert and unification assumptions hold,
then (neglecting two-loop terms)
\begin{mathletters}
\label{toy1}
\begin{equation}
\frac{1}{\alpha_{1}(M_{Z})} = \frac{1}{\alpha_{G}^{0}} + b_{1}t^{0},
\label{toy1a}
\end{equation}
\begin{equation}
\frac{1}{\alpha_{2}(M_{Z})} = \frac{1}{\alpha_{G}^{0}} + b_{2}t^{0},
\label{toy1b}
\end{equation}
\begin{equation}
\frac{1}{\alpha_{s}^{0}(M_{Z})} = \frac{1}{\alpha_{G}^{0}} + b_{3}t^{0}.
\end{equation}
\end{mathletters}
We use (\ref{toy1a}) and (\ref{toy1b}) to define $\alpha_{G}^{0}$ and $t^{0}$
in terms of the (input) $\alpha_{1,2}(M_{Z})$.
We now turn on the $\Delta_{1}$ and $\Delta_{2}$
correction functions and assume that no other corrections
contribute to \rons. The coupling constants are now given by
\begin{mathletters}
\label{toy2}
\begin{equation}
\frac{1}{\alpha_{1}(M_{Z})} = \frac{1}{\alpha_{G}^{0}} + b_{1}t^{0}
+ \Delta_{\alpha_{G}} +b_{1}\Delta_{t} - \Delta_{1},
\end{equation}
\begin{equation}
\frac{1}{\alpha_{2}(M_{Z})} = \frac{1}{\alpha_{G}^{0}} + b_{2}t^{0}
+ \Delta_{\alpha_{G}} +b_{2}\Delta_{t} - \Delta_{2},
\end{equation}
\begin{equation}
\frac{1}{\alpha_{s}(M_{Z})} =
\frac{1}{\alpha_{s}^{0}(M_{Z})} -  \Delta_{\alpha_{s}}
= \frac{1}{\alpha_{G}^{0}} + b_{3}t^{0}
+ \Delta_{\alpha_{G}} +b_{3}\Delta_{t}.
\end{equation}
\end{mathletters}
$\Delta_{\alpha_{G}}$ and $\Delta_{t}$ are determined by the condition
$\Delta_{\alpha_{G}} +b_{i}\Delta_{t} - \Delta_{i} = 0$
for $i$=1, 2,
while (in the present approximation)
$\Delta_{\alpha_{s}}$ is due entirely to the change in
\alphag and $t$, i.e.,
$- \Delta_{\alpha_{s}} = \Delta_{\alpha_{G}} +b_{3}\Delta_{t}$.
Also,
\begin{equation}
\frac{1}{\alpha_{3}(M_{G})} = \frac{1}{\alpha_{G}^{0}}
+ \Delta_{\alpha_{G}},
\label{toy3}
\end{equation}
and the $\alpha_{3}$ term in (\ref{mbmz})
now reads
\begin{equation}
\left(\frac{\alpha_{s}^{0}(M_{Z})}{\alpha_{G}^{0}}\right)^{\frac{8}{9}}\times
\left(1 + \frac{8}{9}[
\alpha_{s}^{0}(M_{Z})\Delta_{\alpha_{s}} +
\alpha_{G}^{0}\Delta_{\alpha_{G}}]
\right).
\label{toy4}
\end{equation}
We thus obtain
(in the toy model)
\begin{equation}
\ro = e^{\frac{8}{9}[(\alpha_{s}(M_{Z}) - \alpha_{G})\Delta_{\alpha_{s}}
 - b_{3}\alpha_{G}\Delta_{t}]}.
\label{toy5}
\end{equation}

In the more general case $\Delta_{3} \neq 0$
and $- \Delta_{\alpha_{s}} = \Delta_{\alpha_{G}} +b_{3}\Delta_{t} -
\Delta_{3}$,
where
$\Delta_{3} = \Delta_{3}^{SUSY}  +\Delta_{3}^{heavy}  +\Delta_{3}^{NRO}$
(for the low-scale threshold, high-scale threshold, and NRO contributions,
respectively).
NRO's ($\Delta_{3}^{NRO}$) modify only the $\alpha_{3}$
value and  not any  RGE coefficients (see Ref. \cite{us} and below)
and can be easily incorporated in our toy model, i.e., (\ref{toy5}) is still
correct if
$- \Delta_{\alpha_{s}} = \Delta_{\alpha_{G}} +b_{3}\Delta_{t} -
\Delta_{3}^{NRO}$.
The high-scale thresholds are more complicated because they not only affect
\als but also change the $\beta$-function coefficient $b_{3}$ and the
coefficient
of the RGE for $y_{b}$ (the $b_{b;3}$) at the various thresholds.
The expression for \ro will be derived below.
Ignoring for now the threshold changes in $b_{b;3}$ the
$\Delta_{3}^{heavy}$ contribution to \ro
$\sim e^{\frac{8}{9}[\alpha_{s}(M_{Z}) - \alpha_{G}]\Delta_{3}^{heavy}}$,
i.e., only the shift in \alsns, which affects the entire $t'$ range in
(\ref{rges}), is relevant. The effect of the change
in $b_{3}$ above the threshold is of second
order because it only affects a small region
of the $t'$ integral.
Similarly, the leading contribution of
$\Delta_{3}^{SUSY}$ is
$\ro \sim e^{\frac{8}{9}[\alpha_{s}(M_{Z}) - \alpha_{3}(\sim \,
TeV)]\Delta_{3}^{SUSY}}$,
which is a second order in small quantities
(because it only affects a small region of the $t'$ integral)
and is therefore negligible.

Hence, the corrections to gauge couplings lead to
\begin{equation}
\ro = e^{\frac{8}{9}[(\alpha_{s}(M_{Z}) - \alpha_{G})\Delta'_{\alpha_{s}}
 - b_{3}\alpha_{G}\Delta_{t}]},
\label{toy6}
\end{equation}
where $-\Delta'_{\alpha_{s}} = -\Delta_{\alpha_{s}} +
\Delta_{3}^{SUSY}
= \Delta_{\alpha_{G}} + b_{3}\Delta_{t} - \Delta_{3}^{NRO} -
\Delta_{3}^{heavy}$
includes all the shifts in \als except those induced by
$\Delta_{3}^{SUSY}$.
The additional corrections associated with the changes
in $b_{b;3}$ at thresholds will be discussed below.

A different complication is due to the non-negligible role of the Yukawa
couplings. $F$ is modified when thresholds are decoupled. In particular,
once the heavy Higgs doublet is decoupled the Yukawa operators and their
evolution
are modified. (Recall that we
assume that we have one heavy ($M_{H} \gg \mz$)
Higgs doublet that decouples with the sparticles, and
another light ($m_{h} \sim \mz$) SM-like doublet that is responsible for all
fermion
masses.)
Also, $h_{\alpha}(M_{G}) > 1$ near either the $h_{t}$ (low-\tgbns) or $h_{b}$
(large-\tgbns)
fixed points,
and the most significant high-scale
effect of correcting $t^{0} \rightarrow t^{0} +\Delta_{t}$ is
due to the large Yukawa couplings and not to the $\alpha_{G}\Delta_{t}$ term.
We will therefore treat high-scale $\Delta_{t}$ effects (\rotns)
separately from $\Delta_{\alpha_{s}}$ and $\Delta_{\alpha_{G}}$
effects.
$\Delta_{\alpha_{s}}$ will include $\Delta_{3}$ contributions which will
be partially cancelled
by decoupling thresholds from both the $\alpha_{3}$
and $y_{b}$ RGE's. Thus,
$\Delta_{\alpha_{s}}$ and $\Delta_{\alpha_{G}}$ effects will be described by
\roalsns, which we derive first. (Using the input
value of $\alpha_{1}$, $\rho_{\alpha_{1}}^{^{-1}} \sim 1$ --
see below.)
We will then consider corrections to $F$ (\ropns). Lastly, we will derive
\rot  and rewrite \ro in a way that reflects the correlations among
\roalsns, \rop and \rot.
We will also comment on the role of the high-scale corrections,
the case of using \als as an input, and on corrections to the $h_{t}/h_{b}$
ratio.

Allowing a complicated threshold structure near \mz (and/or near \mgns) gives a
modified one-loop expression for $m_{b}$,
\begin{equation}
\mb(\mz) = \mtau(\mz)\times
\prod_{i = 1}^{3}\prod_{k = 0}^{n-1}\left(
\frac{\alpha_{i}(\mu^{k})}{\alpha_{i}(\mu^{k + 1})}
\right)^{\frac{b_{b;i}^{k} - b_{\tau;i}^{k}}{2b_{i}^{k}}}
\times F^{-1} \times (1 + \Delta_{F}),
\label{mbol}
\end{equation}
where $k$ runs over the various thresholds,
i.e., $\mu^{0} = \mz$ and $\mu^{n} = \mg$.
$b^{k}$ is the one-loop coefficient of the
respective RGE between
$\mu^{k}$ and $\mu^{k+1}$;
and $\Delta_{F}$ represents the threshold corrections to $F$.
By expanding (\ref{mbol}) around (\ref{mbmz})
(in a similar way to (\ref{toy4}))
and using the results of Ref. \cite{us}
we can obtain an approximate expression
for \rons.
This yields a better insight into the role of the
different correction parameters than purely numerical estimates.

The important effects of the coupling constant uncertainties are in the
$\alpha_{3}$ terms.
$\alpha_{2}$ (in our  approximation\footnote{
Once sparticles are decoupled the degeneracy among various operators is lifted,
e.g., the gaugino - sfermion - fermion coupling
is different from the respective gauge coupling
and the higgsino - sfermion - fermion Yukawa coupling
is different from the Higgs-boson -
fermion - fermion one (see, for example, Chankowski \cite{chak}). In
(\ref{mbol})
we ignored this effect, which is negligible for sparticles and the
Higgs-doublet
below the TeV scale.})
drops out from (\ref{mbol}) and the residual uncertainties from
$\alpha_{1}$ are small when the input value is used.
Recall that our strategy is to use the experimental
values of $\alpha_{1}(M_{Z})$ and $\alpha_{2}(M_{Z})$ to predict
$\alpha_{3}$. The dominant corrections to the \mb prediction
are the uncertainties in \alphag and $t$ due to  $\Delta_{1}$
and $\Delta_{2}$, and the explicit uncertainties
in $\Delta_{3}$ (as was illustrated by our toy model).
 The latter can be divided into low-scale
($\Delta_{3}^{SUSY}$)
and to high-scale ($\Delta_{3}^{heavy} + \Delta_{3}^{NRO}$) contributions.
The low-scale uncertainties have only a small effect
on \mb because they only affect  a small $t'$
range in (\ref{rges})
(see the toy model). High-scale corrections affect the entire $t'$ range.
They modify both \als (high-scale contributions to $\Delta_{3}$ constitute a
part of
$\Delta_{\alpha_{s}}$) and either the $\beta_{3}$ function near \mg
($\Delta_{3}^{heavy}$)
or the $\alpha_{3}(M_{G})$ value
($\Delta_{3}^{NRO}$).
All (high- and low-scale threshold)
corrections to  $\beta_{3}$ affect the $\alpha_{3}$ terms in (\ref{mbol}).

We denote the heavy X and Y vector; color-triplet;
and the adjoint color-octet, $SU(2)$-triplet (and singlet)
superfield thresholds by $M_{V}$, $M_{5}$, and $M_{24}$, respectively.
Some of the high-scale thresholds
are strongly constrained by proton decay,
i.e., in the minimal $SU(5)$ model (which we assume)
$M_{5} \sim M_{G}$
and perturbative consistency constrains $M_{G} \lesssim 3M_{V}$
\cite{an,yan}.
$M_{24} \ll M_{G}$ is possible,
and $\Delta_{t}$ in this scenario can be $\sim + 0.5$ and the constraints on
$M_{5}$ are relaxed (i.e., $M_{5} \gtrsim 0.1\mg$) \cite{yan}.
Also, proton decay constraints can be removed by a simple
modification of the model \cite{ir}.

The sparticles and the Higgs doublet decouple
from the $\alpha_{i}$ RGE
at an effective scale, $M_{i}$, defined in Ref. \cite{us}
(see also Carena et al. \cite{cpw}), i.e.,
\begin{equation}
\sum_{\zeta} \frac{b_{i}^{\zeta}}{(2\pi)}\ln\frac{M_{\zeta}}{M_{Z}} =
\frac{b_{i}^{MSSM} - b_{i}^{SM}}{(2\pi)} \ln\frac{M_{i}}{M_{Z}} \,\, \forss
\label{mi}
\end{equation}
The summation is over all relevant thresholds, i.e.,
sparticles and the heavy Higgs
doublet, and
$b_{i}^{\zeta}$ is the $\zeta$-particle contribution to the respective
$\beta$-function.
$M_{i}$ can be split by a factor of a few. In general, $M_{1}$
grows most significantly with the scalar mass;  $M_{3}$ with the gaugino mass;
and $M_{1}$ and $M_{2}$ grow the same with the higgsino mass;
and   $M_{2} \ll M_{1}$ and/or $M_{2} \ll M_{3}$.
$M_{1}$, $M_{2}$ and $M_{3}$ all appear in $\Delta_{\alpha_{s}}$,
$\Delta_{\alpha_{G}}$ and $\Delta_{t}$. On the other hand,
once either the gluinos or the squarks are decoupled, all squark - gluino
loops are eliminated and $b_{b;3} = b_{b;3}^{SM}$ \cite{chak},
and two other scales of relevance are (in the approximation of
degenerate squark masses)
$\underline{M_{3}} = \min{(M_{gluino},\,M_{squark})}$ and
$\overline{M_{3}} = \max{(M_{gluino},\,M_{squark})}$.
One has $M_{3}^{2} = \overline{M_{3}}\underline{M_{3}}$.

We consider high-scale thresholds and NRO's ($\sim [\als - \alphag]\Delta$),
low-scale  thresholds ($\sim [\als - \alpha_{3}(\sim$ TeV$)]\Delta$),
and corrections to the coupling constant unification predictions for \als
and \alphagns.
We will discuss corrections to $F$ and to $t$ below.
A more detailed  treatment of low-scale effects
will be needed if either some of the spectrum
parameters are better known or if one assumes sparticle thresholds
above the TeV scale.
We will take\footnote{The generalization to $M_{5} < M_{24}$ is straight
forward.
The $M_{V} < M_{G}$ case is much more difficult to describe.
The heavy X and Y supervectors
couple to the $SU_{3}\times SU_{2}\times U_{1}$ Yukawa operators in
a complicated way. However, $M_{V} \gtrsim \frac{1}{3}M_{G}$
and the effects cannot be  large.}
$\mu^{1} = \underline{M_{3}}$,  $\mu^{2} = \overline{M_{3}}$,
$\mu^{3} = M_{24}$ and $\mu^{4} = M_{5}$.
The couplings and coefficients (to be substituted in (\ref{mbol})) read
\begin{mathletters}
\label{expand}
\begin{equation}
\alpha_{3}(\mu^{0}) = \alpha_{s}^{0}(\mz) +
(\alpha_{s}^{0}(\mz))^{2}\Delta_{\alpha_{s}},
\end{equation}
\begin{equation}
(\alpha_{3}(\mu^{1}))^{-1} = (\alpha_{s}^{0}(\mz))^{-1}
-b_{3}^{0}t_{3} - \Delta_{\alpha_{s}},
\end{equation}
\begin{equation}
(\alpha_{3}(\mu^{2}))^{-1} = (\alpha_{s}^{0}(\mz))^{-1}
-b_{3}^{0}t_{3} -b_{3}^{1}\delta t_{3}
- \Delta_{\alpha_{s}},
\end{equation}
\begin{equation}
(\alpha_{3}(\mu^{3}))^{-1} =   (\alpha_{3}(\mu^{2}))^{-1}
- b_{3}^{2}\ln\frac{M_{24}}{\overline{M_{3}}} =
(\alpha_{3}(\mu^{4}))^{-1}
\end{equation}
\begin{equation}
(\alpha_{3}(\mu^{4}))^{-1} =
(\alpha_{G}^{0})^{-1} - b_{3}^{4}t_{5} +\Delta_{\alpha_{G}} - \Delta_{3}^{NRO},
\end{equation}
\begin{equation}
(\alpha_{3}(\mu^{5}))^{-1} = (\alpha_{3}(\mu^{n}))^{-1}
= (\alpha_{G}^{0})^{-1} + \Delta_{\alpha_{G}} - \Delta_{3}^{NRO},
\end{equation}
\begin{equation}
b_{3}^{0} = b_{3}^{SM} = -7, \,b_{3}^{1} = -5,\, b_{3}^{2} = b_{3}^{MSSM} = -3,
b_{3}^{3} = 0, \,b_{3}^{4} = 1,
\end{equation}
\begin{equation}
b_{b;3}^{0} = b_{b;3}^{1} = b_{b;3}^{SM} = -8,\,
b_{b;3}^{2} = b_{b;3}^{3} = b_{b;3}^{4}
= b_{b;3}^{MSSM} = -\frac{16}{3}.
\end{equation}
\end{mathletters}
$t_{3} = \frac{1}{2\pi}\ln \frac{\underline{M_{3}}}{M_{Z}}$,
$\delta t_{3} = \frac{1}{2\pi}\ln \frac{\overline{M_{3}}}{\underline{M_{3}}}$,
and $t_{5} = \frac{1}{2\pi}\ln \frac{M_{5}}{M_{G}}$.
(We replaced $\alpha_{s,G}^{OL}$ by the two-loop
$\alpha_{s,G}^{0}$ which introduces a negligible inconsistency.)
In the $[M_{24}, M_{5}]$ interval we cannot use (\ref{mbol})
because $b_{3}^{3} = 0$,
 and instead we have
$b_{b;3}^{4}\alpha_{3}(\mu^{4})\int_{\mu^{4}}^{\mu^{5}}d\ln\mu'$,
which contributes
$-\frac{4}{3}\frac{\alpha_{G}}{\pi}[\ln\frac{M_{24}}{M_{G}}
- \ln\frac{M_{5}}{M_{G}}]
=  \alpha_{G}[-\frac{8}{9}\Delta^{24}_{3} + \frac{8}{3}t_{5}]$
to $\ln\ro$

We obtain (for the $\alpha_{3}$ terms in (\ref{mbol}))
\begin{equation}
\left( \frac{\alpha_{3}^{OL}(\mz)}{\alpha_{G}^{OL}} \right)^{\frac{8}{9}}
\times \roals,
\label{roalpha1}
\end{equation}
where
\begin{equation}
\roals \equiv
e^{[\frac{8}{9}\alpha_{s}^{0}\Delta_{\alpha_{s}} +
\frac{8}{9}\alpha_{G}^{0}\Delta_{\alpha_{G}} -
\frac{20}{9}\alpha_{s}^{0}t_{3} - \frac{4}{9}\alpha_{s}^{0}\delta t_{3}
- \frac{8}{9}\alpha_{G}^{0}t_{5} - \frac{8}{9}\alpha_{G}^{0}\Delta^{24}_{3}
- \frac{8}{9}\alpha_{G}^{0}\Delta^{NRO}_{3}]}
\label{roalpha}
\end{equation}
($\alpha_{s}^{0}$ is $\alpha_{s}^{0}(M_{Z})$.)

$\alpha_{1}$ (and $\alpha_{2}$)
uncertainties feed into $\Delta_{\alpha_{s}}$,
$\Delta_{\alpha_{G}}$ and $\Delta_{t}$ (we discuss the latter below).
There are also $\rho^{^{-1}}_{\alpha_{1}}$
corrections from thresholds and NRO's analogous to (\ref{roalpha}) from the
$(\alpha_{1}(M_{Z}) /\alpha_{1}(M_{G}))^{\frac{10}{99}}$ factor.
However, these are
negligible ($\lesssim 1\%$ or $\rho^{^{-1}}_{\alpha_{1}} \sim 1$)
when the experimental input value for
$\alpha_{1}(M_{Z})$ is used.
We take in (\ref{mbol}) $\rho^{^{-1}}_{\alpha_{1}} = 1$.
The $\alpha_{1}$ term in (\ref{rges}) does, however,  lead
to a small contribution to the \rot term.

$\Delta_{\alpha_{s}}$, $\Delta_{\alpha_{G}}$, $\Delta_{3}^{NRO}$ and
$\Delta_{3}^{24}$ are
defined in Ref. \cite{us} and are given
in Appendix \ref{sec:app} for completeness.
They involve the low- and high-scale mass parameters introduced above,
as well as the NRO effective strength, $\eta$.
To leading order $\eta$ is the only
NRO free parameter
and  it incorporates the degrees of freedom associated with
the strength, sign, scale, and normalization of the dimension-five operators
$-\frac{1}{2}\frac{\eta}{M_{planck}}Tr(F_{\mu\nu}\Phi F^{\mu\nu})$,
where $F_{\mu\nu}$ is the field strength tensor and $\Phi$ is the adjoint
scalar field. The range $-10 \lesssim \eta \lesssim 10$
suggested in Ref. \cite{us} is constrained only by perturbative consistency
of the analysis.

Threshold corrections also affect the one-loop contribution from the
Yukawa sector, i.e., $F \rightarrow F(1 + \Delta_{F})$, and it is convenient
to define
\begin{equation}
\rop = 1 + \Delta_{F}.
\label{rof1}
\end{equation}
$F^{-1}$ is a correction term,
but it can be as large as a $\sim 30\%$ correction
(which, in fact, is responsible
for the successful \mb prediction in the MSSM),
and, as we shall show, $\Delta_{F} \approx 2\% - 4\%$.
$M_{24} \ll M_{G}$ will not contribute since the adjoint superfield
couples (to one-loop) to the Yukawa operators via its coupling
to the Higgs doublets, which drops out from the ratio. However,
new and large Yukawa couplings will (radiatively) increase $h_{\alpha}(\mu)$
and thus affect the infra-red fixed points
and the perturbative limit; i.e., they affect
$F_{\alpha}$ rather than the ratio $F$. (Such an effect
may shift the $h_{t}$  and $h_{b}$ divergence lines in Figures $2-7$ inwards
towards each other.) New Yukawa operators
(that do contribute to the ratio\footnote{
Their effect can be estimated by
observation of the $SU(5)$ invariant operators, i.e.,
$F_{b}/F_{\tau} \rightarrow 1$ above $M_{5}$, (\ref{f}) is slightly
modified for $M_{5} < M_{G}$, and the divergence
lines  move slightly outwards.})
are also generated if $M_{5} < M_{G}$
(see, for example, Hisano et al. \cite{yan}).
The exact magnitude of
such effects
will be determined by the details of the high-scale Lagrangian.

There are, however, low-scale corrections to $F^{-1}$.
We naively change the Yukawa coupling RGE's
below the heavy Higgs doublet
threshold ($t_{H} = \frac{1}{2\pi}\ln\frac{M_{H}}{M_{Z}}$)
to those which are appropriate given the SM fermion
spectrum with one SM-like Higgs doublet (for
example, see Giveon et al. \cite{yukawaunif}).
We will also neglect (near \mzns)
 $\hb\cos{\beta},\,\,\htau\cos{\beta} \sim 0$.
We obtain
\begin{equation}
\rop = e^{[\frac{1}{2}y_{t} + \frac{3}{4}y_{t}\sin^{2}\beta]t_{H}},
\label{rof}
\end{equation}
where here $y_{t}$ is
taken at \mz (or more correctly, between \mz and $M_{H}$),
and $t_{H} < 0.38$.
\rop increases \mb by slightly diminishing the effect of $F^{-1}$
in (\ref{mbmz}). Note
that the $F$ behavior distinguishes the MSSM, where only \hb gets corrected
(to one-loop) by \htopns,
from the SM where all fermions couple to only
one Higgs doublet, and both \hb and \htau
get corrected.
For $\htop < 1.1$ (as is reasonable at the low-scale)
$\rop \lesssim 1.04$, which is a naive overestimate.
Including a $\hb(\mz) - \htau(\mz)$ ($\lesssim 0.4$) contribution can increase
\rop by less than $\sim 2\%$ (the upper bound is for a large \tgbns).
In most parts of the plane the correction
is  moderate, i.e., $\rop  \lesssim 1.02$
if either $\sin\beta \sim 0$ or $\htop \ll 1$.
Let us stress that this is a somewhat naive description which gets
complicated in many ways. For example, a light $t$-squark and a light
chargino will still couple to the SM-like effective Yukawa operators.
Such effects will have to be accounted for
if and when the spectrum is better known and a refined analysis is required.

Lastly, $t$ (which is determined by $\alpha_{1} - \alpha_{2}$ unification)
can be corrected by either corrections to the coupling constant unification
(see Eq. (\ref{app3})) or by a split between the coupling constant and Yukawa
coupling unification points.
In the latter case, from our definition
of \mgns, $\Delta_{t} < 0$ (and it is reasonable to take $\Delta_{t} \gg  -
1$).
(Effects (e.g., NRO's) that may split $h_{b}(M_{G})$
and $h_{\tau}(M_{G})$ can be also expressed in terms of the split
between the unification points, but then $\Delta_{t}$ has no fixed sign.)
Taking the approximation that $\Delta_{t} \ll t$ so that
$\alpha_{i}(t) \approx \alpha_{i}(t + \Delta_{t})$,
$\htop(t) \approx \htop(t + \Delta_{t})$
we find
\begin{equation}
\rot = e^{(-\frac{1}{2}y_{t} + 2\alpha_{G}^{0})\Delta_{t}},
\label{roti}
\end{equation}
where
$y_{t}$ in (\ref{roti}) is taken at \mg, i.e.,
$y_{t}(M_{G}) = \frac{h_{t}^{2}(M_{G})}{4\pi}$,
and $\hb \approx \htau$ dropped out.
For small values of  $\htop(\mg)$, a longer running time reduces $\hb(\mg)$
(and thus, increases the predicted $\hb(M_{Z})$)
and vice versa. The situation reverses for
$\htop(\mg) \gtrsim \sqrt{16\pi\alphag} \sim \sqrt{2}$.

Collecting our results, we have
\begin{equation}
\ro = \roals \times \rop \times \rot \equiv \rosusy \times \roheavy,
\label{rofinal}
\end{equation}
where
\begin{eqnarray}
\rosusy =
\left(\frac{M_{1}}{M_{Z}}\right)^{[\frac{25}{28}C_{1} +\frac{25}{112}C_{2} +
15C_{8}]}
\times
\left(\frac{M_{2}}{M_{Z}}\right)^{-[\frac{25}{7}C_{1} +\frac{275}{112}C_{2} +
25C_{8}]}
& & \nonumber \\
\times
\left(\frac{\underline{M_{3}}}{M_{Z}}\right)^{[C_{1} + C_{3} - C_{4}]}
\times
\left(\frac{\overline{M_{3}}}{M_{Z}}\right)^{C_{1} + C_{4}}
\times
\left(\frac{M_{H}}{M_{Z}}\right)^{C_{7}}, & &
\label{rozi1}
\end{eqnarray}
and
\begin{eqnarray}
\roheavy =
\left(\frac{M_{V}}{M_{G}}\right)^{-[\frac{3}{7}C_{1} -\frac{37}{14}C_{2} +
24C_{8}]}
\times
\left(\frac{M_{24}}{M_{G}}\right)^{-[\frac{3}{14}C_{1} +\frac{33}{28}C_{2}
-C_{5} + 12C_{8}]}
& & \nonumber \\
\times
\left(\frac{M_{5}}{M_{G}}\right)^{[\frac{9}{14}C_{1} +\frac{1}{28}C_{2}
+ C_{6} +\frac{12}{5}C_{8}]}
\times
\left( 1 + [0.29C_{1} + 0.24C_{2} + C_{8}]\eta\right), & &
\label{rogi1}
\end{eqnarray}
represent the low-scale and high-scale corrections, respectively.
The coefficients $C_{i}$ are defined and estimated in Table \ref{table:t1}.
Note that $C_{1} + C_{3} - C_{4} = 0$, i.e., $\underline{M_{3}}$ drops out.
This is because $\underline{M_{3}}$ is associated with the change in \als
due the threshold, which is a second order effect (see the discussion above).
The $\overline{M_{3}}$ dependence, on the other hand, is due to the
change of the $b_{b;3}$ coefficient and is of first order.
We used $M_{3}^{2} = \overline{M_{3}}\underline{M_{3}}$ and added
$\Delta_{\alpha_{G}}$ and $\Delta_{3}^{NRO}$ $\eta$-terms.

It is instructive to rewrite (using Table \ref{table:t1})
\begin{equation}
\rosusy \approx \left(\frac{M_{1}}{M_{Z}}\right)^{0.02}
\left(\frac{M_{2}}{M_{Z}}\right)^{-0.08}
\left(\frac{\overline{M_{3}}}{M_{Z}}\right)^{0.025}
\left(\frac{M_{H}}{M_{Z}}\right)^{0.01}.
\label{rozi}
\end{equation}
(Different values of $C_{7}$ and $C_{8}$ were averaged.)
If the spectrum were all degenerate at $M_{SUSY}$, then $\rosusy \sim
\left(\frac{M_{SUSY}}{M_{Z}}\right)^{-0.025} \gtrsim 0.94$.
 We can invert the logic
and use (\ref{rozi}) to define an effective scale
that gives \rosusy correctly. For example, in Ref. \cite{us}
we defined an effective  scale parameter, $A_{SUSY}$,
\begin{equation}
25\ln{\frac{M_{1}}{M_{Z}}}
 - 100\ln{\frac{M_{2}}{M_{Z}}}
 + 56\ln{\frac{M_{3}}{M_{Z}}}
 = -19\ln{\frac{A_{SUSY}}{M_{Z}}}.
\label{asusy}
\end{equation}
($A_{SUSY}$ here is $M_{SUSY}$ of Ref. \cite{us},
and we have changed notation in order
to avoid confusion with other definitions of $M_{SUSY}$.)
$A_{SUSY}$ gives correctly the corrections to the \als (or \sthns)
prediction, but does not contain any information on the spectrum -- it
can be as low as a few GeV for sparticles $\gg \mz$.
(See also Carena et al.\cite{cpw}.) Here, we can similarly define
\begin{equation}
\rosusy =
\left(\frac{B_{SUSY}}{M_{Z}}\right)^{[-\frac{19}{28}C_{1} -\frac{250}{112}C_{2}
+ C_{3} +C_{7} - 10C_{8}]}
\sim
\left(\frac{B_{SUSY}}{M_{Z}}\right)^{-0.025}.
\label{bsusy}
\end{equation}
The slightly negative exponent implies in many cases (for non-degenerate
spectra)
$B_{SUSY} \lesssim M_{Z}$
(i.e., $M_{2} < M_{1},\,\overline{M_{3}},\, M_{H}$ usually implies
$\rosusy \gtrsim 1$).
$B_{SUSY} \geq M_{Z}$ for a strongly degenerate spectrum.
For the spectra of Ref. \cite{rr} we find $\rosusy \sim 1$ ($A_{susy} \approx
32,\, 21$ GeV, $B_{SUSY} \approx M_{Z}$).
Taking the limits of heavy gluinos and of a degenerate spectrum
we find $0.94 \lesssim \rosusy \lesssim 1.06$. Away from the limits
$\rosusy \rightarrow 1$.

Similarly to (\ref{rozi}) we can rewrite
\begin{equation}
\roheavy \approx \left(\frac{M_{V}}{M_{G}}\right)^{-0.030}
\left(\frac{M_{24}}{M_{G}}\right)^{-0.004}
\left(\frac{M_{5}}{M_{G}}\right)^{0.015}
\left(1 + 0.007\eta\right).
\label{rogi}
\end{equation}
A scenario in which $M_{24} \ll M_{G}$; $M_{5} \sim (0.1 - 0.5)M_{G}$; $M_{V} =
M_{G}$; and $\eta \approx -10$; would give $\roheavy \approx 0.9$.
This scenario is also consistent with
limits from proton decay \cite{an,yan}.
Furthermore, NRO's contribute only negligibly to $\Delta_{\alpha_{G}}$
and to $\Delta_{t}$
(unless one allows NRO effects to be very large \cite{nro,drees85}).
$M_{24} \ll M_{G}$ on the other hand can increase $t$ significantly,
i.e., $M_{G} \lesssim 5 \times 10^{17}$ GeV
(which is the reason that we can have $M_{5} < M_{G}$).
A large negative $\eta$ maintains an acceptable
value of \als in such a scenario.
Lifting proton decay constraints (e.g., see Ref. \cite{ir}),
 we can have $M_{5} \ll M_{G}$ and $\roheavy \approx 0.8 - 0.9$.
Taking these limits and that of a degenerate spectrum and
a large positive $\eta$ we obtain $0.8 \lesssim \roheavy \lesssim 1.1$

The high-scale corrections to the coupling constant unification emerge as the
leading
contribution to $\ro \neq 1$.
We would like to stress that
$\eta$ is not just a new ad-hoc parameter.
Given the precision to which we know the low-scale
observables, one cannot ignore the likely possibility of unknown physics
at the high-scale where the (supergravity-induced) MSSM breaks down,
and which is  parameterized
in terms of NRO's
(whose form is defined in $SU(5)$ models).
Furthermore, similar corrections may arise in supergravity from
non-minimal (and non-universal) gauge kinetic functions (see, for example,
Ref. \cite{drees85}).
Unfortunately, this,
in turn, introduces some ambiguity in RG calculations
(via high-scale boundary conditions).
It should also be noted that
adding large
representations \cite{fm1,hall2,babu,giudice,mohap},
e.g., $\underline{126}$ of $SO(10)$,
does not introduce (for nearly degenerate heavy components)
large threshold corrections
to \als and $t$. This is because the decoupled heavy components constitute a
nearly
complete representation (which acts equally on all the $b_{i}$'s).
Thus, the threshold corrections
in the minimal model give a good estimate of  \roheavy
(in models with a GUT sector, which are the relevant ones for Yukawa
unification).
A model independent treatment of high-scale threshold effects
on coupling constant unification was given in Ref. \cite{us}.
The heavy Yukawa sectors of different models may affect the
infra-red fixed-points differently.

An arbitrary splitting of the two unification points
induces a $\sim 5\%$ uncertainty.
By combining all the contributions in quadrature (as a guideline only)
we obtain
\begin{equation}
0.80 \lesssim \ro \lesssim 1.15.
\label{rorange}
\end{equation}
$\ro = 1\pm 0.1 $ is thus a reasonable range, and
$\ro = 1\pm 0.15 $ (that we adopted) is somewhat more extreme, but well
within the allowed range. We would like  to stress that all
ranges extracted here are a guideline only.
This range, which is controlled by high-scale corrections, is still valid
when the sparticle spectrum is explicitly calculated (and, e.g., decoupled
numerically).

As we pointed out above, corrections
that either change the prediction for \als or the positive
contribution to (\ref{rges}) from Yukawa terms,
affect the infra-red fixed points and can slightly shift
the corresponding divergence lines in Figures $2 - 7$. They may also
affect the upper bound on \mtpns, and thus they induce a $\sim \pm 10$ GeV
uncertainty to the upper bound value.
However, the corrected range
remains $\gtrsim 200$ GeV
(see also Barger et al. \cite{bbo}). In particular, it
is much higher than the the upper bound  suggested by precision data
(see Eq. (\ref{mtop})).

We would also like to point out that if $\alpha_{1}(M_{Z})$,
$\alpha_{2}(M_{Z})$ and $\alpha_{3}(M_{Z})$ are all used as inputs,
then one arbitrarily adjusts $\Delta_{\alpha_{s}}$ so that
$\alpha_{s}^{0} + (\alpha_{s}^{0})^{2}\Delta_{\alpha_{s}}$ is fixed at some
desired value.
The coupling constants do not unify unless one consistently corrects
$\alpha_{G}$ and $t$ as well.
However, such a procedure is a reasonable approximation if
$\Delta_{3}^{SUSY}$ is small (or is known and corrected for).
In that case one can minimize the
residual uncertainty by calculating $\alpha_{3}(M_{G})$ from
the input value of \als and from \mg (\mg is calculated from
$\alpha_{1}-\alpha_{2}$
unification). Then only \rotns, \rop and $b_{b;3}$ terms contribute
to \ro (i.e., one can  obtain their contribution by
setting $C_{1} = C_{2} \equiv 0$ in (\ref{rozi1}) and (\ref{rogi1})), and the
residual
uncertainty is small.
Some caution is, however, needed. The coupling constant unification constraints
are
not integral in such a procedure (e.g., compare Figures 2 and 5). In
particular,
the correlation between \als and
\mt is not manifest.
A large \mt value implies larger values of
\als or,  alternatively,  very large corrections to coupling constant
unification.
Also, only $\Delta_{1}$, $\Delta_{2}$, $\Delta_{3}^{NRO}$ and
$\Delta_{3}^{heavy}$ can induce first order corrections to \mbns, and thus
can be used to fix \alsns.
(NRO's
renormalize and split $\alpha_{i}(M_{G})$,
and thus honestly modify the boundary conditions.
The simplest  way
to adjust  the \als prediction to a given input value is
by adjusting $\eta$ -- i.e., $\eta\sim -10$ corrects the \als
prediction to $\als \sim 0.11$.)
As was illustrated by our toy model,
$\Delta_{3}^{SUSY}$
contributes to $\Delta_{\alpha_{s}}$ but
does not affect \mb
to first order in small terms.
Thus, unless one knows and corrects for the $\Delta_{3}^{SUSY}$
contribution to the input \als one introduces a significant
theoretical uncertainty.
Lastly, the experimental uncertainty in
\als is large,
and arbitrarily varying \als in that range
is not very instructive.
Nevertheless, it is useful in demonstrating the role of \als in predicting
\mbns,
as we saw in section \ref{sec:plane}.

Finally, the three-Yukawa unification strip (see Figure 2) has uncertainties in
both the \tgb and the \mtp ranges, coming from corrections to the
$h_{b}/h_{\tau}$
and $h_{t}/h_{b}$ ratios, respectively. To one-loop
$(h_{t}/h_{b}) \sim (\alpha_{1}^{OL}/\alpha_{G}^{OL})^{-\frac{1}{33}}
\times(F_{b}/F_{t})$
and any uncertainties in the $\alpha_{1}$ term
are negligible. However, variation of $- 0.5 \lesssim
\Delta_{t} \lesssim 0.5$ generates $\sim \pm 2\%$ ($\sim \pm 8\%$)
correction if $h_{t} \sim h_{b} \sim h_{\tau} \sim 1$
($\sim 2$), i.e., $(\rho_{b}/\rho_{top})_{t} \sim
e^{\frac{1}{2}y_{\alpha}(M_{G})\Delta_{t}}$.
Additional uncertainty of
$0.95 \lesssim (\rho_{b}/\rho_{top})_{F}
\sim e^{(y_{t}(M_{Z}) - \frac{5}{2}y_{b}(M_{Z}))t_{H}} \lesssim 1$
(we assume $\tgb \gg 1$) is associated with the decoupling of the heavy
Higgs doublet. We estimate a $\sim \pm 5 - 10 \%$ uncertainty in the \mtp range
that corresponds to three-Yukawa unification.

\section{Conclusions}
\label{sec:con}
Grand unified theories typically predict $h_{b} = h_{\tau}$ at \mgns, and
contain non-fundamental Higgs representations.
These distinguish such models from some other realizations of the
MSSM, e.g., string-inspired GUT-like models. Above, we
explicitly embedded the MSSM in a minimal $SU(5)$
model, and concluded that such a model is constrained to a small
area of the parameter space.
We showed that corrections to a two-loop calculation of
the bottom mass (when assuming grand unification)
are manifested in various ways. Parametrizing those corrections,
we were able to relate them to the correction parameters
identified in Ref. \cite{us}, and to study their magnitude and behavior in some
detail.
The theoretical uncertainty in the bottom mass prediction
is typically $\lesssim 15\%$.
We thus took (given the ambiguities in the extraction of \mb
from experiment)  $0.85\mbo(5\,\gev) < 4.45$ GeV
as a (conservative) constraint.
Requiring this, as well as requiring perturbative Yukawa couplings up to \mg
and identifying the coupling constant and the (third-family)
Yukawa coupling unification points,
we found that the range $2.7 \lesssim \tgb \lesssim 40$ is excluded
(as well as $\mtp \gtrsim  215$ GeV),
and that, in agreement with other authors, the allowed area in the $\mtp -
\tgb$
plane is described by low- and high-\tgb branches (where the
former saturates the $h_{t}$ fixed-point line).
The separation between the two branches is determined by the correction factor.
Requiring all three (third-family) Yukawa couplings to meet constrains
$160\,\gev\, \lesssim \mtp$ and requires a large \tgbns.
We demonstrated that the allowed parameter space grows for lower (input) values
of
\alsns, but that the MSSM prefers higher values.
We further argued that the \sth quadratic dependence on \mtp cannot
be ignored as
it correlates the \als prediction with \mtpns,
and thus affects the \mtp dependence
of the \mb prediction, as well as the
the upper bound on \mtp  and the range of \mtp
for which intermediate values of \tgb are allowed.
Finally, we expect the above observations
and radiative breaking of $SU(2) \otimes U(1)$
to have
mutual implications,
and suggest
that the above constraint is still valid
in  a calculation
in which the sparticle spectrum, and therefore \rosusyns, is calculated
explicitly. (The larger uncertainty in the calculation comes
from the unification-scale physics rather than from the details of the
sparticle spectrum.)
Our hope is that a careful study of various correction terms will
eventually result in
reliable constraints on the
MSSM parameter space,
and in a way that can distinguish different realizations of the MSSM.
Here we have showed (in agreement with others) that by
measuring \tgb one can exclude
simple (and some extended) GUT structures at the high-scale.

\acknowledgments
This work was supported by the Department of Energy
Grant No. DE-AC02-76-ERO-3071.

\appendix
\section{The Correction Functions}
\label{sec:app}
For completeness, we give the correction
functions  to the coupling constant unification
(in the minimal $SU(5)$ MSSM).
For more details, see Ref. \cite{us}.
Corrections that depend on \mtp or the conversion to the $\overline{DR}$
scheme are included in the numerical procedure, and are not quoted below.
All the parameters are defined above (see section \ref{sec:corrections}).

\begin{eqnarray}
{28\pi}\Delta_{\alpha_{s}} =
  -12\ln{\frac{M_{V}}{M_{G}}}
  - 6\ln{\frac{M_{24}}{M_{G}}}
  + 18\ln{\frac{M_{5}}{M_{G}}} & & \nonumber \\
   + 25\ln{\frac{M_{1}}{M_{Z}}}
   -100\ln{\frac{M_{2}}{M_{Z}}}
   + 56\ln{\frac{M_{3}}{M_{Z}}}
   +8.00\eta. & &
\label{app1}
\end{eqnarray}

\begin{eqnarray}
 -{336\pi}\Delta_{\alpha_{G}} =
  +888\ln{\frac{M_{V}}{M_{G}}}
  - 396\ln{\frac{M_{24}}{M_{G}}}
  + 12\ln{\frac{M_{5}}{M_{G}}} & & \nonumber \\
   + 75\ln{\frac{M_{1}}{M_{Z}}}
   -825\ln{\frac{M_{2}}{M_{Z}}}
   +50.0\eta. & &
\label{app2}
\end{eqnarray}

\begin{eqnarray}
\frac{336\pi}{5}\Delta_{t} =
 -24\ln{\frac{M_{V}}{M_{G}}}
 - 12\ln{\frac{M_{24}}{M_{G}}}
 + \frac{12}{5}\ln{\frac{M_{5}}{M_{G}}} & & \nonumber \\
 + 15\ln{\frac{M_{1}}{M_{Z}}}
 -25\ln{\frac{M_{2}}{M_{Z}}}
 +1.0\eta. & &
\label{app3}
\end{eqnarray}

\begin{equation}
\Delta_{3}^{24} = \frac{3}{2\pi}\ln\frac{M_{24}}{M_{G}}.
\label{app4}
\end{equation}

\begin{equation}
\Delta_{3}^{NRO} \approx 0.03\eta.
\label{app5}
\end{equation}

\begin{figure}
\caption{The predicted strong coupling at the $Z$-pole, \alsns, for different
values of the $t$-quark pole mass, \mtpns, and of the two Higgs doublet
expectation value ratio, \tgbns. $\hbg = \htaug$ is assumed.
\mtp (in GeV) is indicated on the right-hand-side above the relevant line.}
\end{figure}

\begin{figure}
\caption{The $\mtp - \tgb$ plane is divided into five different regions.
Two areas (low- and high-\tgb branches) are consistent with perturbative
two-Yukawa unification ($\hbg = \htaug$)
and with $0.85\mbo(5\,\gev) < 4.45$ GeV. Between the two branches the $b$-quark
mass is too high. For a too low (high) \tgbns, \htop (\hbns)
diverges. The strip where all three (third-family) Yukawa couplings
unify intersects
the allowed high-\tgb branch and is indicated as well (dash-dot line).
Corrections to the $h_{t}/h_{b}$ ratio induce
a $\sim \pm 5\%$ (vertical) uncertainty in the \mtp range that corresponds
to each of the points in the three-Yukawa unification strip.
\alsns, \alphagns,  and the unification scale used in the calculation are the
ones
predicted by the MSSM coupling constant unification, and
are sensitive to the $t$-quark pole mass, \mtp  (see Figure 1). The \mtp range
suggested by the electroweak data is indicated (dashed lines) for comparison.
\mtp is in GeV.}
\end{figure}
\begin{figure}
\caption{The low-\tgb branch of Figure 2 is shown in greater detail. The lines
corresponding to $\ro = 1$ (thick) and $\htg = 2$ (dashed) are indicated.
To the left of the allowed branch one obtains $\htg > 3$.}
\end{figure}

\begin{figure}
\caption{The same as Figure 2, except  the constraint is replaced
 with the more restrictive one, $0.85\mbo(4.45\,\gev) < 4.45$ GeV.
The allowed \tgb range is reduced by $\sim 0.03 - 0.10$ for the low-\tgb
branch (the effect is hardly seen in the figure) and by
$\sim 3 - 4$ for the high-\tgb branch (where the corresponding range
for 0.85$\mbo(5 \, \gev) < 4.45$ GeV is indicated -- dashed line -- for
comparison).}
\end{figure}

\begin{figure}
\caption{The area in the $\mtp-\tgb$ plane which is consistent with
perturbative two-Yukawa unification and with $0.85\mbo(5\,\gev) < 4.45$ GeV
assuming $\als = 0.11$. The unification scale and \alphag
used in the calculation are those predicted by $\alpha_{1} - \alpha_{2}$
unification. We chose $\ro =0.85$ for comparison with Figures $2 -3$.
\mtp is in GeV.}
\end{figure}

\begin{figure}
\caption{The same as Figure 5, except $\als = 0.12$.}
\end{figure}

\begin{figure}
\caption{The same as Figure 5, except $\als = 0.13$.}
\end{figure}

\begin{table}
\caption{The coefficients $C_{i}$ are defined and estimated using $\sth =
0.2324$,
$\alpha_{s}^{0} = 0.125$, and $\alpha_{G}^{0} = 0.040$.}
\label{table:t1}
\begin{tabular}{c c c c}
 &
definition&
estimate&
comments\\
\hline\hline
$C_{1}$&
$\frac{8}{9}\frac{\alpha_{s}^{0}(M_{Z})}{\pi}$&
$+0.035$ &
\\
\hline
$C_{2}$&
$-\frac{8}{9} \frac{\alpha_{G}^{0}}{\pi}$ &
$-0.011$ &
\\
\hline
$C_{3}$&
$-\frac{10}{9} \frac{\alpha_{s}^{0}(M_{Z})}{\pi}$ &
$-0.044$ &
\\
\hline
$C_{4}$&
$-\frac{2}{9} \frac{\alpha_{s}^{0}(M_{Z})}{\pi}$ &
$-0.009$ &
\\
\hline
$C_{5}$&
$-\frac{4}{3} \frac{\alpha_{G}^{0}}{\pi}$ &
$-0.017$ &
$M_{24} < M_{5}$
\\
\hline
$C_{6}$ &
$-\frac{4}{9} \frac{\alpha_{G}^{0}}{\pi}$ &
$-0.006$ &
$M_{24} < M_{5}$
\\
\hline
$C_{7}$&
$\frac{2 + 3\sin\beta}{8}\frac{y_{t}(M_{Z})}{\pi}$ &
\begin{tabular}{c}
$+ 0.010$\\
$+ 0.008$\\
$+ 0.019$\\
\end{tabular}&
\begin{tabular}{c}
$h_{t} \sim 0.8$, $\beta \sim \frac{\pi}{2}$ \\
$h_{t} \sim 1.1$, $\beta \sim 0 $\\
$h_{t} \sim 1.1$, $\beta \sim \frac{\pi}{2}$ \\
\end{tabular} \\
\hline
$C_{8}$&
$\frac{5}{672} \frac{-y_{t}(M_{G}) + 4\alpha_{G}^{0}}{\pi}$ &
\begin{tabular}{c}
$+ 0.0002$\\
$- 0.0013$\\
\end{tabular}&
\begin{tabular}{c}
$h_{t} \sim 1$ \\
$h_{t} \sim 3$ \\
\end{tabular} \\

\end{tabular}
\end{table}


\begin{thebibliography}{50}
\bibitem{cpw} \CPW; \OP. (Note the different notation, e.g., their $T_{i}$
is our $M_{i}$.)
\bibitem{florida} For a recent review of various observables, SM RGE's,
$\beta$- and
   $\gamma$- functions, see \Arasonetal.
   Note that some values used there, i.e., those for \als and $m_{\tau}$, are
   different from ours, and that the authors follow Narison \cite{nar} in their
   definition of \mbns.
\end{thebibliography}
\end{document}